# Producing Useful Work in a Cycle by Absorbing Heat from a Single Thermal Reservoir: An Investigation on a Locally Nonchaotic Energy Barrier

Yu Qiao,[1,2,*] Zhaoru Shang[1]

[1] *Program of Materials Science and Engineering, University of California – San Diego, La Jolla, CA 92093, U.S.A.*

[2] *Department of Structural Engineering, University of California – San Diego, La Jolla, CA 92093-0085, U.S.A.*

[*] Corresponding author (Email: yqiao@ucsd.edu; phone: 858-534-3388)

**Abstract:** In the current research, we investigate the concept of spontaneously nonequilibrium dimension (SND), and show that a SND-based system can break the second law of thermodynamics. The main characteristic of the SND is the inherent nonequilibrium particle crossing ratio. A locally nonchaotic energy barrier is employed to form the model system. On the one hand, when the barrier width is much smaller than the mean free path of the particles, the system cannot reach thermodynamic equilibrium; on the other hand, the nonequilibrium particle distribution allows for production of useful work in a cycle by absorbing heat from a single thermal reservoir. Such system performance is demonstrated by a Monte Carlo simulation. It should be attributed to the unbalanced cross-influence of the thermally correlated thermodynamic forces, incompatible with the conventional framework of statistical mechanics. No Maxwell's demon is involved. Similar effects may be achieved by a number of variants, e.g., when the barrier is switchable or there are distributed nonchaotic traps.

## 1. Introduction: Concept and Hypothesis

Randomness is a fundamental concept in statistical mechanics [1,2]. An ergodic and chaotic system always tends to reach thermodynamic equilibrium, while a nonergodic or nonchaotic system may not [e.g., 3-5]. The latter is often analyzed in the framework of nonequilibrium stochastic thermodynamics [6].

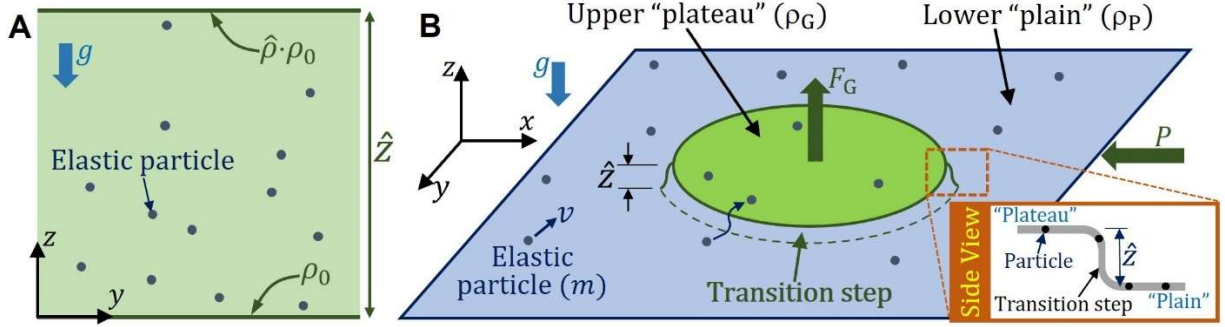

**Figure 1. (A)** In a vertical plane in a gravitational field ($g$), if the plane height ($\hat{z}$) is much less than the mean free path of the particles ($\lambda_F$), the steady-state particle density does not follow the Maxwell-Boltzmann distribution ($\hat{\rho} \neq \delta_0$). **(B)** The billiard-type model system, in which elastic particles randomly move in the upper "plateau" and the lower "plain", across the narrow transition step. When $\hat{z} \ll \lambda_F$, the particle trajectories in the transition step tend to be nonchaotic, and the steady-state particle density ratio ($\hat{\rho} = \rho_G/\rho_P$) is inherently nonequilibrium ($\hat{\rho} \neq \delta_0$).

One example of the nonequilibrium steady state is shown in Figure 1(A), in which elastic particles randomly move in a vertical plane in a uniform gravitational field ($g$). The $y$ dimension is infinitely large. The plane height is denoted by $\hat{z}$. The upper boundary ($z = \hat{z}$) and the lower boundary ($z = 0$) are diffusive walls at constant temperature ($T$), where the reflected particle velocity follows the two-dimensional (2D) Maxwell-Boltzmann distribution, $p(v) = \frac{mv}{\bar{K}} \cdot \exp\left(-\frac{mv^2}{2\bar{K}}\right)$ [7], with $v$ being the particle velocity, $m$ the particle mass, $\bar{K} = k_B T$ the average particle kinetic energy ($K$), and $k_B$ the Boltzmann constant. If the mean free path of the particles ($\lambda_F$) is much smaller than the plane height ($\hat{z}$), the system is ergodic and chaotic. At $z = \hat{z}$, the local particle density, $\hat{\rho}\rho_0$, is equal to $\delta_0 \rho_0$ [8], where $\hat{\rho}$ is the particle density ratio, $\rho_0$ is the local particle density at the bottom of the plane ($z = 0$), $\delta_0 = e^{-\beta m g \hat{z}}$ is the Boltzmann factor, and $\beta = \frac{1}{k_B T}$. If $\lambda_F \gg \hat{z}$, the characteristic travel time ($\bar{t}_T = 2\hat{z}/\bar{v}_z$) is much shorter than the characteristic time of particle collision ($\bar{t}_C = \lambda_F/\bar{v}$), where $\bar{v}_z$ is the average $v_z$, $v_z$ is the z-component of particle velocity, and $\bar{v}$ is the average particle velocity. As the particle-particle interaction is negligible, the particle trajectories tend to be nonchaotic. The probability for a particle to climb from $z = 0$ to $z = \hat{z}$ is mainly determined by its z-direction kinetic energy ($K_z = mv_z^2/2$), relatively unrelated to the particle movement in the $y$ direction. The particle density ratio can be assessed as $\hat{\rho} = \int_{\sqrt{2g\hat{z}}}^{\infty} p_z(v_z) dv_z = \delta_0^2$, where $p_z(v_z) = \sqrt{\frac{2m}{\pi K}} \exp\left(-\frac{mv_z^2}{2K}\right)$ is the one-dimensional Maxwell-Boltzmann distribution of $v_z$. In general, the nonchaotic $\hat{\rho}$ is less than $\delta_0$.

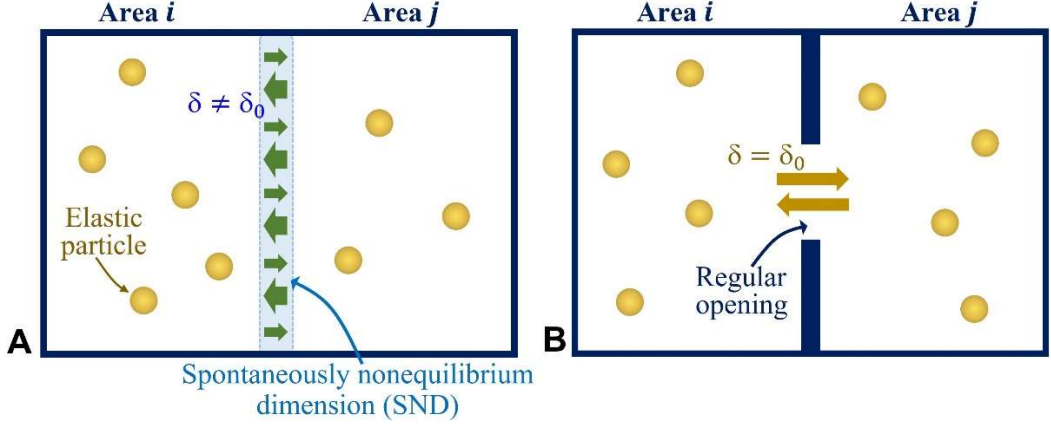

**Figure 2. (A)** In an elastic-particle system that contains a spontaneously nonequilibrium dimension (SND), the steady-state distribution of the particles is intrinsically in a non-Boltzmann form; i.e., $\hat{\rho} \neq \delta_0$. **(B)** In an ergodic and chaotic system, at thermodynamic equilibrium, $\hat{\rho} = \delta_0$.

Figure 2(A) depicts two chaotic and ergodic areas ($i$ and $j$) that are separated by a locally nonchaotic and/or nonergodic barrier. A large number of particles randomly move in the system. Like the example in Figure 1(A), the key characteristic of the barrier is that its crossing ratio, $\delta = \delta_{ji}/\delta_{ij}$, is inherently nonequilibrium, where $\delta_{ij}$ and $\delta_{ji}$ are the probabilities for the particles to cross the barrier from area $i$ to $j$ and from area $j$ to $i$, respectively. For an elastic-particle system, it means that $\delta \neq \delta_0$, where $\delta_0 = e^{-\beta \cdot \Delta E}$ is the Boltzmann factor, and $\Delta E$ is the potential difference between the two areas. Hereafter, such a barrier will be referred to as a spontaneously nonequilibrium dimension (SND). Without any detailed knowledge of the system microstate, the SND offers a mechanism to reach a nonequilibrium steady state, i.e., $\hat{\rho} \neq \delta_0$, where $\hat{\rho} = \rho_i/\rho_j$, and $\rho_i$ and $\rho_j$ are the steady-state particle densities in areas $i$ and $j$, respectively. In comparison, Figure 2(B) shows the same two chaotic and ergodic areas connected through a regular opening. Because across the opening $\delta = \delta_0$, at thermodynamic equilibrium $\hat{\rho} = \delta_0$ [8].

We hypothesize that the behavior of a SND-based system can contradict the second law of thermodynamics. With the SND, useful work may be produced in a cycle by absorbing heat from a single thermal reservoir.

In the current study, we investigate a SND-based system that satisfies the following six conditions: 1) The system is divided by the SND into two chaotic and ergodic areas ($i$ and $j$). 2) Areas $i$ and $j$ are dominated by two different thermodynamic forces ($F_i$ and $F_j$), respectively. 3) $F_i$ and $F_j$ are thermally correlated; that is, $F'_{ij}$ and $F'_{ji}$ are associated with particle redistribution, where $F'_{ij} = \frac{\partial F_i}{\partial \tilde{x}_j}$ and $F'_{ji} = \frac{\partial F_j}{\partial \tilde{x}_i}$ are the cross-influence of $F_i$ and $F_j$, and $\tilde{x}_i$ and $\tilde{x}_j$ are the conjugate variables of $F_i$ and $F_j$, respectively. In other words, when $\tilde{x}_i$ varies, $F_j$ would change, not directly caused by the work of $\tilde{x}_i$, but resulting from the particle diffusion; likewise, $\tilde{x}_j$ also indirectly affects $F_i$. Furthermore, the operations of $F_i$ and $F_j$ are 4) reversible and 5) independent of each other, and 6) do not rely on temperature variation.

In Section 2 below, we define the billiard-type model system that contains a locally nonchaotic energy barrier. In Section 3, a Monte Carlo simulation is performed. The numerical result indicates that spontaneously, the steady-state particle distribution across the barrier is nonequilibrium (Section 3.2). In Section 4, we show that the nonequilibrium particle distribution is incompatible with the second law of thermodynamics, which is also demonstrated by the simulation data in Section 3.3. Extended discussion is given in Section 5.

## 2. The Model System

### 2.1 Setup of the billiard-type model system

Figure 1(B) shows a billiard-type model system of the SND. A large number of elastic particles randomly move in the horizontal $x$–$y$ dimension. A uniform gravitational field ($g$) is along the out-of-plane direction, $-z$. The central area is higher, which will be referred to as the "plateau". The surrounding lower area will be referred to as the "plain". The plateau and the plain are separated by a vertical transition step. The step imposes an energy barrier to the particle movement from the plain to the plateau.

The total particle number ($N$) is constant. The height of the transition step ($\hat{z}$) can be changed by raising or lowering the plateau. The plain area ($A_P$) can be adjusted by moving the outer system boundary. The area of the plateau ($A_G$) does not vary. The particle densities on the plain and the plateau are defined as $\rho_P = N_P/A_P$ and $\rho_G = N_G/A_G$, respectively; $N_P$ and $N_G$ are the particle numbers on the plain and the plateau, respectively. The thermodynamic forces under investigation are the support force of the plateau ($F_G$) and the in-plane pressure of the plain ($P$), with the conjugate variables being $\hat{z}$ and $-A_P$, respectively. It is assumed that i) the particle motion is frictionless; ii) the changes of $\hat{z}$ and $A_P$ are reversible; iii) the transition step is smooth, i.e., as the particles move across it, no energy is dissipated; and iv) the system is immersed in a thermal bath, with a fixed average particle kinetic energy ($\bar{K} = k_B T$).

If $\hat{z} \gg \lambda_F$, the system is ergodic and chaotic. If $\hat{z} \ll \lambda_F$, the transition step becomes a locally nonchaotic SND. The model system meets the six conditions listed in the second to last paragraph in the introductory section: The plateau and the plain are large, in which the particle movement is chaotic and ergodic; they are separated by the energy barrier of the transition step; they are dominated by $F_G$ and $P$, respectively; $F_G$ has a different governing equation from $P$; a variation in $A_P$ would change $F_G$, and a variation of $\hat{z}$ would change $P$; the cross-influence of $F_G$ and $P$ is associated with a particle redistribution across the transition step, accompanied by a heat exchange with the environment; $A_P$ and $\hat{z}$ can be adjusted reversibly and independently, and the system temperature is maintained constant.

The in-plane pressure of the plain ($P$), the plain area ($A_P$), the support force of the plateau ($F_G$), and the plateau height ($\hat{z}$) can be measured from the surface of the system. In the following discussion, the system state will be defined by these macroscopic variables.

2.2 Governing equations

At the outer boundary of the plain, the particles collide with the system wall. The change in particle momentum is balanced by the reaction force of the wall, based on which we can obtain the ideal-gas law for the in-plane pressure of the plain [7]

$$P \cdot A_P = N_P \cdot \bar{K} \qquad (1)$$

The support force of the plateau ($F_G$) has two components: the weight of the particles on the plateau $F_{Gg} = mgN_G$, and the centrifugal force ($F_{Gc}$) caused by the particles changing direction in the transition step. Denote the average time for a particle to change its direction by $\tilde{t}_p$. At the steady state, during $\tilde{t}_p$, on average $\rho_G \bar{v}_x \tilde{t}_p L_G$ particles cross the plateau-step border, where $\bar{v}_x = \sqrt{2k_B T/(\pi m)}$ is the average velocity in one direction, and $L_G$ is the plateau circumference. The average centrifugal force of a particle is on the scale of $m\bar{v}_x/\tilde{t}_p$. Thus, $F_{Gc}$ may be approximately estimated as $m\rho_G \bar{v}_x^2 L_G$, and $F_G = F_{Gg} - F_{Gc} \approx m(gA_G - \bar{v}_x^2 L_G)\rho_G$. For a circular plateau with the diameter of $D_G$, $A_G = \pi D_G^2/4$ and $L_G = \pi D_G$. Define the characteristic plateau height as $z_0 = \frac{\bar{K}}{mg}$. When $D_G \gg z_0$, $F_{Gg} \gg F_{Gc}$, so that $F_G$ can be simplified as $F_G = F_{Gg}$, i.e.,

$$F_G = mgN_G \tag{2}$$

## 3. Monte Carlo Simulation

3.1 Setup of the Monte Carlo simulation

Nonequilibrium stochastic processes are often analyzed through Monte Carlo (MC) simulation. The computer program of our MC simulation for the model system is based on [9] and is available in the Supplementary Material [10]. The setup is shown in Figure 3.

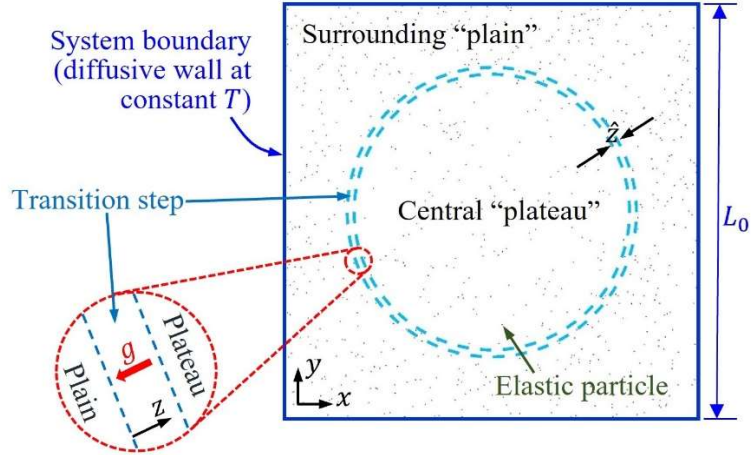

**Figure 3.** The Monte Carlo simulation of 800 billiard-like particles freely moving in the surrounding "plain" and the central "plateau", across the transition step. The dashed circles indicate the boundaries of the transition step with the outer plain and the central plateau.

The simulation box represents the surface of particle movement. The outer boundary is diffusive, where the reflected particle velocity follows the 2D Maxwell-Boltzmann distribution, $p(v)$. The system is divided into two areas by a narrow circular band: the surrounding plain and the central plateau. The circular band is the transition step, in which the local dimension is denoted by $z$, along the radius direction toward the center. No long-range force is applied on the particles in the plain and the plateau. The particles in the transition step are subject to a constant force, $mg$, along $-z$. The width of the transition step is denoted by $\hat{z}$.

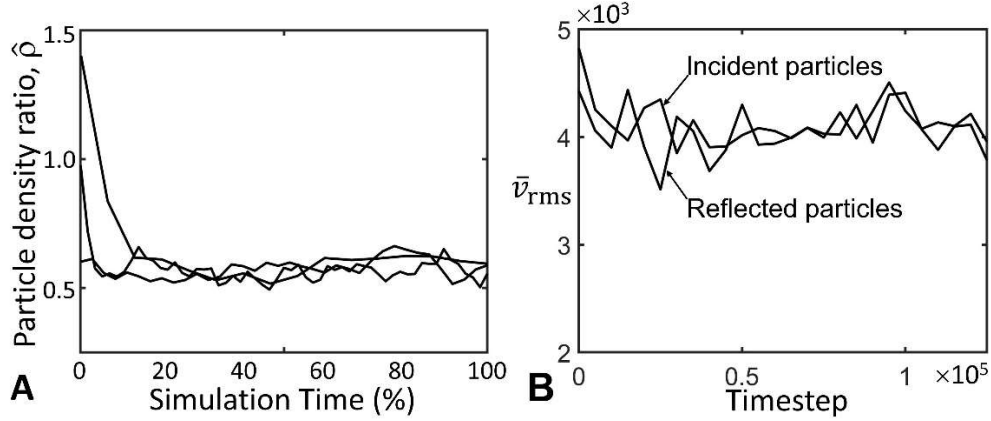

**Figure 4. (A)** Three time profiles of $\hat{\rho}$, showing that the initial $\hat{\rho}$ does not affect the steady-state $\hat{\rho}$. The parameter setting is similar to Case R3 in Table 1, except that the gravitational acceleration is two times smaller. The initial particle distribution is adjusted, so that the initial $\hat{\rho}$ is 0.6, 1.0, or 1.4; the total simulation time is $0.6 \times 10^5$, $0.9 \times 10^5$, or $0.35 \times 10^5$ timesteps, respectively. **(B)** Typical time profiles of the root mean square velocity ($\bar{v}_{\text{rms}}$) of the incident particles and the reflected particles at the system boundary. The parameter setting is the same as Case R3 in Table 1.

The simulation setup is scalable; an example of the unit system can be based on Å, fs, g/mol, and K. The total particle number is $N = 800$; temperature $T = 1000$; the average particle kinetic energy $\bar{K} = 8.3142 \times 10^{-4}$; the particle mass $m = 1$; the simulation timestep is $\Delta t_0 = 1$; the normalization factor of the particle diameter ($d$) is $d_0 = 1.6$; the plateau area is $A_G = 1.256 \times 10^5$; the normalization factor of $\hat{z}$ is $\hat{z}_0 = 20$; the normalization factor of $g$ is $g^* = 4.245 \times 10^{14} g_n$, where $g_n = 9.8 \times 10^{-20}$ is the standard gravity in the sample unit system.

For each simulation case, at each timestep, the particle numbers on the plain ($N_P$) and on the plateau ($N_G$) are counted. The average particle density ratio $\hat{\rho} = \rho_G/\rho_P$ is computed for every $1 \sim 4 \times 10^3$ timesteps. The support force of the plateau ($F_G$) is calculated from Equation (2), averaged for every $1 \sim 2.5 \times 10^3$ timesteps. The in-plane pressure of the plain ($P$) is obtained as $(\Sigma_b \Delta \bar{p}_x)/(\Delta \bar{t} \cdot 4L_0)$, where $\Sigma_b$ indicates summation for all the particle-boundary collisions in every $\Delta \bar{t} = 4 \sim 8 \times 10^3$ timesteps, $4L_0$ is the perimeter of the simulation box, and $\Delta \bar{p}_x$ is the change in particle momentum in the normal direction. Initially, the particles are evenly placed on the plain and the plateau; the particle velocity follows $p(v)$ and the direction is random. Each simulation case is continued until the steady state has been reached. The initial $\hat{\rho}$ does not affect the steady-state particle density ratio (see Figure 4A). At the steady state, there is no overall heat exchange with the environment (see Figure 4B).

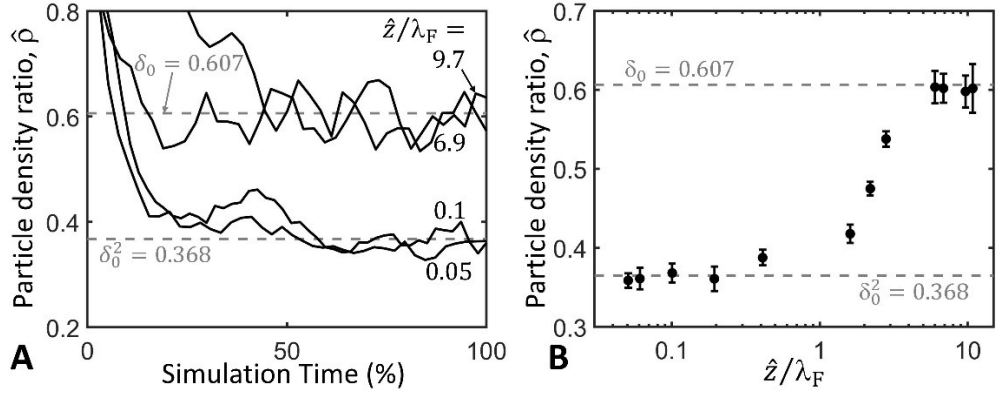

**Figure 5. (A)** Typical time profiles of $\hat{\rho}$. From bottom to top, the four curves correspond to Cases R1, R3, R10, and R11 in Table 1, respectively; the simulation times are $0.5\times10^5$, $0.6\times10^5$, $0.47\times10^5$, and $0.45\times10^5$ timesteps, respectively. **(B)** The steady-state $\hat{\rho}$ as a function of the $\hat{z}/\lambda_F$ ratio. From left to right, the twelve data points correspond to Cases R1 to R12 in Table 1, respectively. It can be seen that $\hat{z}/\lambda_F$ significantly affects $\hat{\rho}$: $\hat{\rho}$ converges to the Boltzmann factor ($\delta_0 = e^{-\beta mg\hat{z}}$) when $\hat{z}/\lambda_F$ is relatively large (i.e., when the system is chaotic), and converges to $\delta_0^2$ when $\hat{z}/\lambda_F$ is relatively small (i.e., when the transition step is locally nonchaotic).

3.2 Simulation result: Nonequilibrium particle distribution

Figure 5(A) gives typical time profiles of $\hat{\rho}$. Figure 5(B) shows the steady-state $\hat{\rho}$ as a function of the $\hat{z}/\lambda_F$ ratio. The error bars are calculated as the 90%-confidence interval, $\pm 1.645\, s_t/\sqrt{n_t}$, where $s_t$ is the standard deviation and $n_t$ is the number of data points. For different simulation cases, the nominal mean free path of the particles, $\lambda_F \approx \frac{A_P+A_G}{\sqrt{8}Nd}$ [11], is adjusted by changing the particle diameter ($d$) and the plain area ($A_P$). The value of $\hat{z}$ is also varied; $g$ is controlled to keep $mg\hat{z}/\overline{K} = 0.5$, so that the Boltzmann factor $\delta_0 = \exp(-mg\hat{z}/\overline{K})$ remains constant 0.607. Table 1 lists the parameter settings.

When $\hat{z}/\lambda_F$ is much smaller than 1, the transition step is locally nonchaotic, and the steady-state $\hat{\rho}$ is close to $\delta_0^2 = 0.368$; i.e., the particle distribution is nonequilibrium. It is consistent with the analysis of Figure 1(A), and should be attributed to that in the energy barrier, without extensive particle collision, the particle motion along $z$ is mainly dependent on $K_z$. The trend is clear that the steady-state $\hat{\rho}$ increases with the $\hat{z}/\lambda_F$ ratio, especially in the range of $\hat{z}/\lambda_F$ from 1 to 4. When $\hat{z}/\lambda_F$ is relatively large, the transition step is chaotic, and the steady-state $\hat{\rho}$ converges to $\delta_0 = 0.607$, i.e., the system reaches thermodynamic equilibrium.

Table 1 The simulation cases in Figure 5

| Case Number | $\hat{z}/\lambda_F$ | $A_P/A_G$ | $d/d_0$ | $\hat{z}/\hat{z}_0$ | $g/g^*$ | Calculated $\hat{\alpha}$ |
|---|---|---|---|---|---|---|
| R1 | 0.0507 | 1.7628 | 0.6250 | 0.4000 | 1.2500 | 2.0500±0.0260 |
| R2 | 0.0608 | 1.7628 | 0.7500 | 0.4000 | 1.2500 | 2.0372±0.0386 |
| R3 | 0.1006 | 1.7628 | 1.0000 | 0.5000 | 1.0000 | 1.9983±0.0326 |
| R4 | 0.1939 | 1.7628 | 1.0000 | 1.0000 | 0.5000 | 2.0383±0.0441 |
| R5 | 0.4110 | 1.7628 | 1.2500 | 1.8000 | 0.2778 | 1.8956±0.0242 |
| R6 | 1.6069 | 1.7628 | 2.2375 | 5.0000 | 0.1000 | 1.7455±0.0238 |
| R7 | 2.1966 | 1.7628 | 2.8125 | 5.7500 | 0.0870 | 1.4893±0.0188 |
| R8 | 2.8077 | 1.7628 | 3.3750 | 6.4500 | 0.0775 | 1.2405±0.0113 |
| R9 | 6.0179 | 3.1831 | 7.5000 | 10.000 | 0.0500 | 1.0100±0.0242 |
| R10 | 6.8617 | 4.7746 | 8.7500 | 15.000 | 0.0333 | 1.0150±0.0218 |
| R11 | 9.6756 | 6.3662 | 12.500 | 25.000 | 0.0200 | 1.0287±0.0245 |
| R12 | 10.827 | 7.9577 | 15.000 | 30.000 | 0.0167 | 1.0157±0.0259 |

3.3 Simulation result: Production of useful work in an isothermal cycle

Figure 6 shows an isothermal operation cycle. The parameters are based on Case R3 in Table 1, with $\hat{z}/\lambda_F$ around 0.1 and $z_0 = \overline{K}/(mg) = \hat{z}_0$. The error bars are computed as the 90%-confidence interval, $\pm 1.645\, s_t/\sqrt{n_t}$.

At State I, $\hat{z}/z_0 = 0.25$ and $A_P/A_G = 0.8879$. From State I to II, $A_P$ is constant and $\hat{z}/z_0$ increases to 0.5. As $\hat{z}$ rises, less particles are on the upper plateau, so that $F_G$ decreases while $P$ becomes larger. From State II to III, $\hat{z}$ is constant and $A_P/A_G$ expands to 1.7637. Since the particle density is reduced, both $F_G$ and $P$ are smaller. From State III to IV, $A_P$ does not vary and $\hat{z}/z_0$ is lowered back to 0.25. Because the energy barrier of the transition step is less, $F_G$ increases and $P$ decreases. Finally, the system returns to State I, and the densification of the particles leads to the increase in both $F_G$ and $P$. For each simulated system state, Table 2 lists $\hat{z}/z_0$, $A_P/A_G$, as well as the steady-state $F_G$, $P$, and $\hat{\rho}$. In addition to States I, II, III, and IV, there are 12 intermediate states, marked as I-a, I-b, etc.

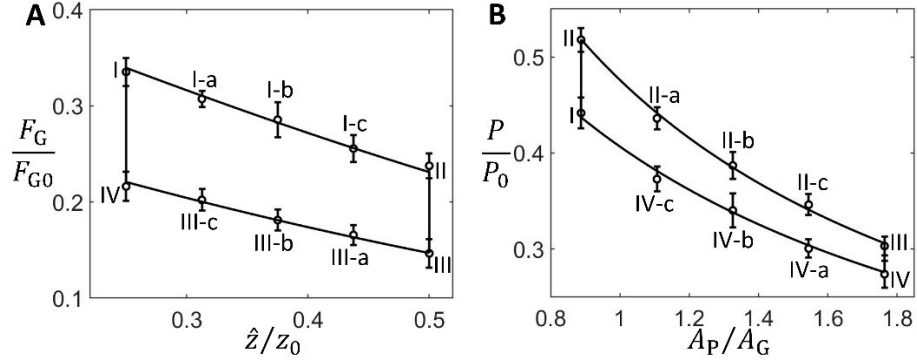

**Figure 6.** An isothermal cycle based on Case R3 in Table 1. The system evolves from State I to II, III, IV, and back to State I. The solid regression curves are calculated from Equations (5) and (6). The work produced by $P$ ($W_P$) is greater than the work consumed by $F_G$ ($W_G$): $W_P/W_G = 1.71$. The normalization factors are $F_{G0} = mgN$, $z_0 = \bar{K}/(mg)$, and $P_0 = N\bar{K}/A_G$.

Table 2 State evolution of the isothermal cycle in Figure 6

| System state | $\hat{z}/z_0$ | $A_P/A_G$ | $F_G/F_{G0}$ | $P/P_0$ | $\hat{\rho}$ |
|---|---|---|---|---|---|
| I | 0.2500 | 0.8879 | 0.3354 | 0.4417 | 0.5932 |
| I-a | 0.3125 | 0.8879 | 0.3073 | 0.4453 | 0.5066 |
| I-b | 0.3750 | 0.8879 | 0.2857 | 0.4536 | 0.4758 |
| I-c | 0.4375 | 0.8879 | 0.2556 | 0.4620 | 0.4056 |
| II | 0.5000 | 0.8879 | 0.2377 | 0.5179 | 0.3556 |
| II-a | 0.5000 | 1.1069 | 0.2164 | 0.4361 | 0.3859 |
| II-b | 0.5000 | 1.3258 | 0.1920 | 0.3871 | 0.3924 |
| II-c | 0.5000 | 1.5448 | 0.1652 | 0.3462 | 0.3754 |
| III | 0.5000 | 1.7637 | 0.1466 | 0.3032 | 0.3683 |
| III-a | 0.4375 | 1.7637 | 0.1657 | 0.2817 | 0.4404 |
| III-b | 0.3750 | 1.7637 | 0.1812 | 0.2803 | 0.4476 |
| III-c | 0.3125 | 1.7637 | 0.2022 | 0.2789 | 0.5522 |
| IV | 0.2500 | 1.7637 | 0.2163 | 0.2739 | 0.6340 |
| IV-a | 0.2500 | 1.5448 | 0.2347 | 0.3006 | 0.6235 |
| IV-b | 0.2500 | 1.3258 | 0.2610 | 0.3403 | 0.6155 |
| IV-c | 0.2500 | 1.1069 | 0.2832 | 0.3729 | 0.5840 |

From State I to II (through States I-a, I-b, and I-c) and from State III to IV (through States III-a, III-b, III-c), the plain-to-plateau area ratio ($A_P/A_G$) remains constant. As $\hat{z}$ varies, the particle density ratio ($\hat{\rho}$) changes. Since both the plateau size ($D_G$) and the plain size ($L_0$) are much larger than $\hat{z}$, $N \approx N_P + N_G$. In accordance with Equation (2), the support force of the plateau is

$$F_G = \frac{mgN}{1+\frac{A_P}{\hat{\rho}A_G}}. \qquad (3)$$

Under the condition of local equilibrium [8], $\hat{\rho}$ can be generally written as

$$\hat{\rho} = \exp\left(-\hat{\alpha}\frac{mg\hat{z}}{\overline{K}}\right) \tag{4}$$

where $\hat{\alpha}$ is a parameter independent of location; it reflects the effect of the local nonchaoticity. If $\hat{\alpha} = 1$, the right-hand side of Equation (4) is reduced to the Boltzmann factor. When $\hat{z}/\lambda_F$ is small, $\hat{\alpha} > 1$ (see Figure 5). Combination of Equations (3) and (4) gives

$$F_G = \frac{mgN}{1+\frac{A_P}{A_G}\exp\left(\hat{\alpha}\frac{mg\hat{z}}{\overline{K}}\right)}, \tag{5}$$

where $\hat{\alpha}$ is the only adjustable parameter for data fitting; all the other parameters are known. When $\hat{\alpha}$ is set to 2.28, Equation (5) agrees well with the MC simulation data of the $F_G - \hat{z}$ relationship from State I to II (the upper curve in Figure 6A). When $\hat{\alpha}$ is set to 2.06, Equation (5) agrees well with the simulation data from State III to IV (the lower curve in Figure 6A).

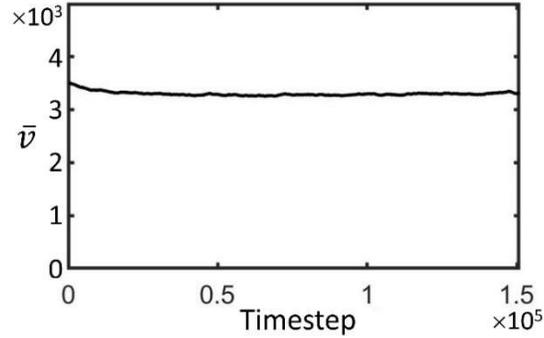

**Figure 7.** Typical time profile of the average particle velocity ($\bar{v}$) in the reference system. Except for $\hat{z} = 0$, the other parameters are the same as in Case R3 in Table 1.

From State II to III (through II-a, II-b, and II-c) and from State IV to I (through IV-a, IV-b, and IV-c), $\hat{z}$ is kept constant while the plain area ($A_P$) changes. Based on Equation (1), the simulated in-plane pressure of the plain can be expressed as

$$P = \frac{N\overline{K}_0}{A_P+\bar{\rho}A_G} \tag{6}$$

where $\bar{\rho}$ is the average particle density ratio and $\overline{K}_0 = \frac{1}{2}m\bar{v}^2 = 5.4459 \times 10^{-4}$ is calculated from the reference system with $\hat{z} = 0$ (see Figure 7). The upper and the lower solid curves in Figure 6(B) are from Equation (6). For the upper curve, $\bar{\rho}$ is 0.3755, the average $\hat{\rho}$ of States II, II-a, II-b, II-c, and III (Table 2); for the lower curve, $\bar{\rho}$ is 0.6100, the average $\hat{\rho}$ of States IV, IV-a, IV-b, IV-c, and I (Table 2). It can be seen that Equation (6) is in agreement with the simulation result of the $P - A_P$ relationship.

The $F_G - \hat{z}$ loop consumes work $W_G = -\left|\int_I^{II} F_G d\hat{z}\right| + \int_{III}^{IV} F_G d\hat{z} = 20.24\overline{K}$, calculated as the area enclosed by the upper and the lower $F_G - \hat{z}$ curves in Figure 6(A). The integral bounds are the system states. The $P - A_P$ loop produces work $W_P = \int_{II}^{III} P dA_P - \left|\int_{IV}^{I} P dA_P\right| = 34.61\overline{K}$, calculated as the area in between the upper and the lower $P - A_P$ curves in Figure 6(B). The ratio between $W_P$ and $W_G$ is $W_P/W_G = 1.71$. After a complete cycle, the overall work production is $W_{tot} = W_P - W_G = 14.37\overline{K}$, about 41.5% of $W_P$.

## 4. Conflict with the Second Law of Thermodynamics

In this section, we discuss how $W_P > W_G$ can be plausible for the model system. It is attributed to the nonequilibrium steady-state particle distribution ($\hat{\rho} \neq \delta_0$).

<u>4.1 Cross-influence of the thermodynamic forces</u>

Denote the Helmholtz free energy by $\mathcal{A}$. If the model system is at thermodynamic equilibrium, because $F_i = \frac{\partial \mathcal{A}}{\partial \tilde{x}_i}$ and $F_j = \frac{\partial \mathcal{A}}{\partial \tilde{x}_j}$ [8] (also see Section 5.4 below), $\frac{\partial F_i}{\partial \tilde{x}_j} = \frac{\partial^2 \mathcal{A}}{\partial \tilde{x}_i \partial \tilde{x}_j} = \frac{\partial F_j}{\partial \tilde{x}_i}$, i.e.,

$$F'_{ij} = F'_{ji} \tag{7}$$

That is, the cross-influence of $F_i$ and $F_j$ is symmetric.

Equation (7) represents the Kelvin-Planck statement of the second law of thermodynamics, i.e., no useful work can be produced in a cycle by absorbing heat from a single thermal reservoir [12], as illustrated by the isothermal cycle in Figure 8. With an arbitrarily small increment of $d\tilde{x}_i$ or $d\tilde{x}_j$, the work that $F_i$ or $F_j$ does to the system is $F_i \cdot d\tilde{x}_i$ or $F_j \cdot d\tilde{x}_j$, respectively. From State I to II, $\tilde{x}_j$ is constant; $\tilde{x}_i$ increases by $d\tilde{x}_i$, and the cross-influence causes $F_j$ to vary by $F'_{ji} d\tilde{x}_i$. From State II to III, $\tilde{x}_i$ is constant; $\tilde{x}_j$ decreases by $-d\tilde{x}_j$, and the cross-influence causes $F_i$ to vary by $-F'_{ij} d\tilde{x}_i$. From State III to IV, $\tilde{x}_j$ does not vary; $\tilde{x}_i$ decreases by $-d\tilde{x}_i$ and correspondingly, $F_j$ varies by $-F'_{ji} d\tilde{x}_i$. From State IV to I, the system returns to the initial state; $\tilde{x}_j$ rises by $d\tilde{x}_j$, and $F_i$ is changed by $F'_{ij} d\tilde{x}_j$. The overall work done by $F_j$ to the environment is $W_j = F'_{ji} d\tilde{x}_i d\tilde{x}_j$; the overall work done by $F_i$ is $W_i = -F'_{ij} d\tilde{x}_j d\tilde{x}_i$. The $F_i - \tilde{x}_i$ loop results in a heat loss $\Delta q_i$; the $F_j -$ $\tilde{x}_j$ loop causes a heat absorption $\Delta q_j$. After a complete cycle, the second law of thermodynamics demands that $\Delta q_i + \Delta q_j = 0$ and $W_i + W_j = 0$ [13], both of which lead to Equation (7).

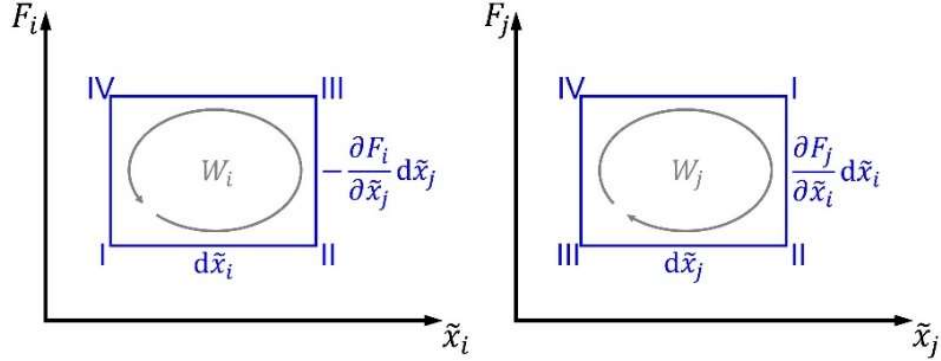

**Figure 8.** The Kelvin-Planck statement of the second law of thermodynamics dictates that the cross-influence of $F_i$ and $F_j$ must be symmetric, i.e., $F'_{ij} = F'_{ji}$. The system operates in an isothermal cycle. Indexes I-IV indicate the system states; $d\tilde{x}_i$ and $d\tilde{x}_j$ are arbitrarily small. Without losing generality, the two loops are drawn for negative $F'_{ij}$ and $F'_{ji}$.

4.2 Nonequilibrium particle distribution

In Figure 1(B), we investigate $P$ and $F_G$, for which Equation (7) becomes

$$\frac{\partial P}{\partial \hat{z}} = -\frac{\partial F_G}{\partial A_P} \tag{8}$$

When the plateau size is large and the step size is small, the system is governed by Equations (1) and (2), and $N \approx N_G + N_P$. Under this condition, Equation (8) can be rewritten as $\frac{z_0}{A_P}\frac{\partial(\overline{N}_G)}{\partial \hat{z}} = \frac{\partial(\overline{N}_G)}{\partial A_P}$, to satisfy which $\overline{N}_G$ needs to be $f\left[\frac{A_P}{A_G}\exp\left(\frac{\hat{z}}{z_0}\right)\right]$, where $\overline{N}_G = N_G/N$ and $f$ is a certain function. When $\hat{z} = 0$, $\overline{N}_G = f\left(\frac{A_P}{A_G}\right) = \frac{A_G}{A_G+A_P}$. Therefore, $f(\bullet) = \frac{1}{1+\bullet}$, so that the solution of Equation (8) is

$$\overline{N}_G = \frac{1}{1+\frac{A_P}{A_G}\exp\left(\frac{\hat{z}}{z_0}\right)} = \frac{\tilde{A}_G}{\tilde{A}_G+A_P} \tag{9}$$

where $\tilde{A}_G = A_G\exp\left(-\frac{\hat{z}}{z_0}\right) = A_G e^{-\beta mg\hat{z}}$. That is, the balance of the cross-influence of $P$ and $F_G$ (Equation 8) is equivalent to the Maxwell-Boltzmann distribution of the particle density,

$$\hat{\rho} = \frac{\rho_G}{\rho_P} = e^{-\beta mg\hat{z}} \tag{10}$$

Equation (10) is derived from the system governing equations (Equations 1 and 2), not through the Lagrange multiplier method [7]. It is required by the Kelvin-Planck statement of the second law of thermodynamics (Equation 8), no matter whether the transition step is chaotic or nonchaotic.

However, the MC simulation clearly indicates that $\hat{\rho}$ is significantly influenced by the $\hat{z}/\lambda_F$ ratio, as demonstrated in Figure 5. In Table 1, $\hat{\alpha}$ is calculated from Equation (4), by using the numerical result of $\hat{\rho}$ in Figure 5(B). Because Equation (10) is the solution of Equation (8), if $\hat{\rho} \neq \delta_0$ ($\hat{\alpha} \neq 1$), Equation (8) cannot be satisfied, i.e., $\frac{\partial P}{\partial \hat{z}} \neq -\frac{\partial F_G}{\partial A_P}$; in other words,

$$F'_{ij} \neq F'_{ji} \tag{11}$$

The mechanism of $\hat{\rho} \neq \delta_0$ is related to the crossing ratio of the locally nonchaotic SND, $\delta$. Similar to the example in Figure 1(A), when $\hat{z}/\lambda_F \ll 1$, the relevant kinetic energy for the particles to overcome the gravitational energy barrier is mostly determined by $K_z$, on average less than $\overline{K}$ by a factor of 2. Consequently, $\delta \to \delta_0^2$.

### 4.3 Production of useful work in the isothermal cycle

Inequality (11) causes the overall work production shown in Figure 6. Based on $\rho_G = \hat{\rho} \cdot \rho_P$ and $N = N_G + N_P$, we have $N_P = N/[1 + \hat{\rho} A_G/A_P]$ and $N_G = N/[1 + A_P/(\hat{\rho} A_G)]$. According to Equations (1) and (4), $W_P = \overline{K}\left[\int_{II}^{III} \frac{N_P}{A_P} dA_P - \int_{I}^{IV} \frac{N_P}{A_P} dA_P\right] = N\overline{K} \cdot \ln\left(\frac{A_{Pu}+\hat{\rho}_{II}A_G}{A_{Pl}+\hat{\rho}_{II}A_G} \frac{A_{Pl}+\hat{\rho}_I A_G}{A_{Pu}+\hat{\rho}_I A_G}\right)$, where $A_{Pl}$ and $A_{Pu}$ are the compressed plain area and the expanded plain area, respectively; $\hat{\rho}_I$ and $\hat{\rho}_{II}$ are the particle density ratios associated with the lowered plateau and the raised plateau, respectively; the integral bounds are the system states. Likewise, based on Equations (2) and (4), $W_G = mg\left[\int_{I}^{II} N_G \cdot d\hat{z} - \int_{III}^{IV} N_G \cdot d\hat{z}\right] = \frac{N\overline{K}}{\hat{\alpha}} \cdot \ln\left(\frac{A_{Pl}+\hat{\rho}_I A_G}{A_{Pl}+\hat{\rho}_{II}A_G} \frac{A_{Pu}+\hat{\rho}_{II}A_G}{A_{Pu}+\hat{\rho}_I A_G}\right)$. Hence, $\frac{W_P}{W_G} = \hat{\alpha}$. If $\hat{z}/\lambda_F \gg 1$, the transition step is chaotic and $\hat{\rho} = \delta_0$ (i.e., $\hat{\alpha} = 1$), so that $W_P/W_G = 1$, in agreement with Equation (8). If $\hat{z}/\lambda_F \ll 1$, the transition step is a locally nonchaotic SND and $\delta \to \delta_0^2$ (i.e., $\hat{\alpha} \to 2$), which leads to

$$\frac{W_P}{W_G} = 2, \tag{12}$$

contradicting the second law of thermodynamics.

In Section 3.3, the calculated $W_P/W_G$ ratio is 1.71, smaller than the ideal-case scenario of Equation (12). The difference may be attributed to the occasional particle collisions in the transition step in the MC simulation, the finite size and the curvature of the transition step, the imperfect simplification of Equation (2), the boundary effect, the local heterogeneity and anisotropy, and the occupied area of the particles. For Equation (5), these factors also render the numerical result of $\hat{\alpha}$ different from 2.

### 4.4 Counterexample of intrinsically symmetric $F'_{ij}$ and $F'_{ji}$

It is worth noting that the operation of the nonequilibrium system does not necessarily cause a contradiction to the second law of thermodynamics (Equation 7). A counterexample may be formed if the conditions listed in the second to last paragraph in the introductory section are violated, e.g., when both thermodynamic forces are chosen as the in-plane pressures: $F_i = P$ and $F_j = P_G$, where $P_G$ is the in-plane pressure of the plateau. It breaks Condition 2, since the plain and the plateau are governed by the same equation of state. With such $F_i$ and $F_j$, Equation (7) becomes $\frac{\partial P}{\partial A_G} = \frac{\partial P_G}{\partial A_P}$. Based on Equation (1), it can be rewritten as $-A_G \frac{\partial N_G}{\partial A_G} = A_P \frac{\partial N_G}{\partial A_P}$, which has the general solution of $N_G = f^*\left(\frac{A_P}{A_G}\right)$, with $f^*$ being a certain function. Because $N_G = N/\left(1 + \frac{1}{\hat{\rho}}\frac{A_P}{A_G}\right)$, the solution is always satisfied. It suggests that the cross-influence of $P$ and $P_G$ is intrinsically symmetric, regardless of whether $\hat{\rho}$ obeys the Maxwell-Boltzmann distribution or a non-Boltzmann distribution.

## 5. Discussion

### 5.1 Variants of the SND-based system

There can be a variety of different types and configurations of SND. For example, recently, inspired and encouraged by the current numerical study on locally nonchaotic energy barrier, we experimentally investigated a locally nonchaotic entropy barrier [14]. A series of tests were carried out on a microporous polyamide membrane one-sidedly surface-grafted with dodecyl chains. The organic chains behaved as molecular-sized outward-swinging gates. The measurement data demonstrated entropy decrease without an energetic penalty. The second law of thermodynamics was generalized, based on the principle of maximum entropy.

Figure 1(B) represents an ideal-gas system in a thermal bath. It has various variants, such as Figure 9(A). The right-hand side of the plateau boundary is connected to the lower plain by a narrow vertical step, and the left-hand side is connected through a wide ramp. The height of the transition step ($\hat{z}$) is much less than $\lambda_F$, while the ramp width ($\hat{L}$) is much larger than $\lambda_F$. Therefore, the particle behavior is nonchaotic in the transition step, but chaotic in the ramp. When the step and the ramp are alternately closed and reopened (i.e., the plain-plateau border is switchable), the system may shift between the equilibrium state and the nonequilibrium steady state. When the ramp is closed by a frictionless sliding door and the transition step is open, similar to Figure 1(B), the particle density ratio ($\hat{\rho}$) is less than the Boltzmann factor ($\delta_0$). As the plain area expands from $A_{Pl}$ to $A_{Pu}$, the system does work $W_+$ to the environment. Then, the ramp is opened and the transition step is closed by another frictionless sliding door. Since the entire system becomes ergodic and chaotic, $\hat{\rho}$ equals to $\delta_0$. As the plain area is reduced from $A_{Pu}$ back to $A_{Pl}$, because the equilibrium $N_P$ is less than the nonequilibrium $N_P$, the work that the environment does to the system ($W_-$) is smaller than $W_+$. After such a cycle, useful work ($W_{tot}$) is generated, equal to the thermal energy absorbed from the environment. In accordance with $N_P = NA_P/(A_P + \hat{\rho}A_G)$, $W_{tot} = W_+ - W_- = \frac{N}{\beta} \cdot \tilde{\Delta}\left(\ln \frac{A_{Pu} + \hat{\rho}A_G}{A_{Pl} + \hat{\rho}A_G}\right)$, where $\tilde{\Delta}(\bullet)$ indicates the difference between the equilibrium state and the nonequilibrium state.

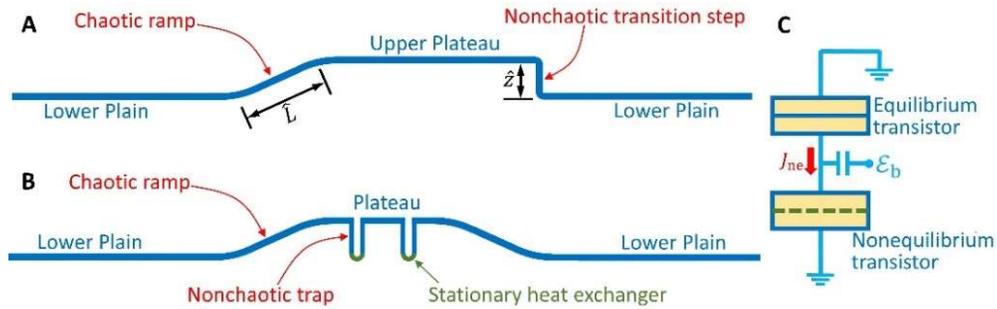

**Figure 9. (A)** Side view of a variant of the billiard-type model system. The plateau-plain border is switchable. **(B)** Another variant of the model system. The plateau and the plain are connected through a wide ramp. There are nonchaotic traps distributed on the plateau. **(C)** A locally nonequilibrium transistor couple.

Figure 9(B) is another variant of Figure 1(B). The entire plateau-plain border is a chaotic wide ramp, while there are a number of locally nonchaotic vertical traps distributed on the plateau. Particles randomly move into and out of the traps. The bottoms of the traps are perfect heat exchangers at constant $T$, and do not move, so that the trap depth changes with the plateau height ($\hat{z}$). The trapped particles have no contribution to the support force of the plateau, $F_G$. Useful work may be produced if $\hat{z}$ and $A_P$ are alternately adjusted in an isothermal cycle similar to Figure 6.

5.2 Comparison to Maxwell's demon

In essence, the locally nonchaotic energy barrier in Figure 1(B) is a directional "filter". On the one hand, from the plateau to the plain, it allows all the particles to pass; on the other hand, from the plain to the plateau, it blocks the particles that are slow in the normal direction ($v_z < \sqrt{2g\hat{z}}$ at the plain-step border). No specific microstate is directly observed. The system operation does not involve the physical nature of information. The working mechanism relies on the SND-induced nonequilibrium steady state, resulting from the unforced thermal motion of the particles. This is fundamentally different from Maxwell's demon [15-18].

The classical Maxwell's demon is nonequilibrium, but not spontaneous [19]. Its overall effect is somewhat similar to that of the SND, i.e., to render $\delta \neq \delta_0$. Yet, it is subject to the energetic penalty associated with the information flow [20,21]. For example, in the on-chip Maxwell's demon enabled by the single-electron transistors (SET) [17], at each operation step, the electron transition always obeys the Fermi-Dirac distribution. The final nonequilibrium state is caused by the external interruption, which consumes work to obtain and process the detailed knowledge of system microstates.

Autonomous Maxwell's demon (AMD) is spontaneous, but not nonequilibrium. For instance, Feynman's ratchet [22] may be viewed as an AMD-type device. A set of vanes are connected to a set of ratchet and pawl, immersed in a gas container. The vanes and the ratchet undergo rotational Brownian movements, overcoming the energy barrier of the pawl ($\Delta E_p$) with the same Boltzmann factor, $e^{-\beta \cdot \Delta E_p}$. The time-average behaviors of them counterbalance each other. The concept of AMD has been experimentally demonstrated by a single-electron refrigerator [18], which consists of two SET. One SET plays the role of the ratchet in Feynman's model (the "autonomous demon"); the other SET plays the role of the vanes; the capacitive coupling between the two SET corresponds to the energy barrier of the pawl. There is no analog of the bias voltage ($\mathcal{E}_b$) in Feynman's original setup. A counterpart of $\mathcal{E}_b$ may be added by attaching a constant-torque spiral spring on the vanes; the spring uses the stored elastic energy to drive the system to rotate in the forward direction of the ratchet. Likewise, in the single-electron refrigerator, electric energy is consumed from the power supply of the bias voltage, as electrons move across the system to decrease the local entropy. As all the electron-tunneling events follow the Fermi-Dirac distribution, mere structural asymmetry does not result in any conflict with the second law of thermodynamics.

In a SET-based system, in order to reduce entropy without an energetic penalty, one of the transistors needs to be inherently nonequilibrium (i.e., its intrinsic probability density function of electron transition is a non-Fermi distribution). Figure 9(C) depicts a nonequilibrium transistor connected to a regular transistor. They are subjected to opposite bias voltages ($\mathcal{E}_b$). Due to the different transition rates, a net flux of conduction electrons ($J_{ne}$) could be generated. The electron behaviors at the two transistors are uncorrelated; therefore, the system is not a variant of Maxwell's demon. The overall work done by $\mathcal{E}_b$ is zero. Somewhat similarly, in the asymmetric setup in Figure 9(A), if both sides of the plain-plateau border are open, an ordered macroscopic particle flow may be spontaneously formed.

5.3 Nonequilibrium maximum of entropy

As discussed in [14], the SND can apply additional constraints on the probability of system microstates. Consequently, when entropy ($S$) is maximized, it reaches the nonequilibrium maximum ($S_{ne}$), lower than the global maximum at thermodynamic equilibrium ($S_{eq}$). In other words, with the SND, the maximum possible entropy of steady state ($S_Q$) is less than $S_{eq}$.

In Figure 1(B), the constraints from the locally nonchaotic energy barrier may be expressed as $p_\mu = \kappa_{\mu\nu} p_\nu$, where $p_\mu$ and $p_\nu$ are the probabilities of the $\mu$-th and the $\nu$-th microstates, respectively; $\kappa_{\mu\nu} = \delta^{N_{\mu\nu}}$; $N_{\mu\nu}$ is the difference in $N_G$ between the $\mu$-th and the $\nu$-th microstates. It represents the effect of the "deterministic" particle transition across the step. Since $\delta$ and $N_{\mu\nu}$ are affected by the step height and the plain size, $S_{ne}$ is a function of $\hat{z}$ and $A_P$. As $\hat{z}$ and $A_P$ are changed alternately in the isothermal cycle in Figure 6, $S_Q$ shifts between a higher $S_{ne}$ and a lower $S_{ne}$, causing the production of useful work. This process cannot be described by the classical statements of the second law of thermodynamics, but is consistent with the generalized form [14]:

$$S \to S_Q \tag{13}$$

That is, in an isolated system, $S$ has the tendency to converge toward $S_Q$, not $S_{eq}$.

### 5.4 Limitation of the Helmholtz free energy

Because the steady state of the model system is nonequilibrium, the thermodynamic forces cannot be directly calculated from the Helmholtz free energy, $\mathcal{A}$ [23], i.e.,

$$F_i \neq \frac{\partial \mathcal{A}}{\partial \tilde{x}_i} \text{ and } F_j \neq \frac{\partial \mathcal{A}}{\partial \tilde{x}_j}. \tag{14}$$

It renders the derivation of Equation (7) irrelevant, allowing for Inequality (11).

In Figure 1(B), as $\hat{z}$ is much smaller than the plateau size and the plain size, $N \approx N_P + N_G$, and the direct contribution of the narrow step to $S$ is negligible. Hence,

$$S = S_1 + S_2 \tag{15}$$

where $S_1 = N_P k_B \left(\ln \frac{eA_P}{N_P} + \sigma\right)$ is the entropy of the plain, $S_2 = N_G k_B \left(\ln \frac{eA_G}{N_G} + \sigma\right)$ is the entropy of the plateau, and $\sigma$ is a function of $m\overline{K}$; the expressions of $S_1$ and $S_2$ are based on the equation of entropy of idea gas [8]. Accordingly, the Helmholtz free energy is

$$\mathcal{A} = U - TS = N_G mg\hat{z} - k_B T \left(N_P \ln \frac{A_P}{A_G} - N_P \ln N_P - N_G \ln N_G\right) + \tilde{\sigma} \tag{16}$$

where $U$ is the internal energy and $\tilde{\sigma} = -Nk_B T(\ln A_G + \sigma)$. If $\hat{\rho} = \delta_0$, Equation (16) is reduced to $-k_B T \cdot \ln Z$, with $Z$ being the partition function. With $\hat{\rho} = \frac{N_G}{A_G}/\frac{N_P}{A_P} = \exp\left(-\hat{\alpha} \frac{mg\hat{z}}{k_B T}\right)$ (Equation 4), Equation (16) leads to $-\frac{\partial \mathcal{A}}{\partial A_P} = \frac{N_P k_B T}{A_P} + \frac{\partial N_G}{\partial A_P}(\hat{\alpha} - 1)mg\hat{z}$ and $\frac{\partial \mathcal{A}}{\partial \hat{z}} = mgN_G - \frac{\partial N_G}{\partial \hat{z}}(\hat{\alpha} - 1)mg\hat{z}$. By replacing $N_G$ by $\frac{N}{1+A_P/(\hat{\rho}A_G)}$, we have

$$P = -\frac{\partial \mathcal{A}}{\partial A_P} + (\hat{\alpha} - 1)N \frac{mg\hat{z}A_G\hat{\rho}}{(A_P + A_G\hat{\rho})^2} \tag{17}$$

$$F_G = \frac{\partial \mathcal{A}}{\partial \hat{z}} - (\hat{\alpha} - 1)\hat{\alpha}N \frac{A_P}{z_0} \frac{mg\hat{z}A_G\hat{\rho}}{(A_P + A_G\hat{\rho})^2} \tag{18}$$

where $P = N_P k_B T/A_P$ and $F_G = mgN_G$ are given by Equations (1) and (2), respectively. Clearly, to satisfy the classical relationships of $P = -\frac{\partial \mathcal{A}}{\partial A_P}$ and $F_G = \frac{\partial \mathcal{A}}{\partial \hat{z}}$, for a nontrivial case with a nonzero $mg\hat{z}A_G\hat{\rho}$, the system must be at thermodynamic equilibrium (i.e., $\hat{\alpha} = 1$). For the nonequilibrium model system, because $\hat{\alpha} > 1$, $-\frac{\partial \mathcal{A}}{\partial A_P}$ predicts a lower pressure than $P$, and $\frac{\partial \mathcal{A}}{\partial \hat{z}}$ predicts a larger force than $F_G$. It reflects that the particle density ratio across the locally nonchaotic step ($\hat{\rho}$) is less than the Boltzmann factor, $\delta_0$. Equations (17) and (18) confirm that $\frac{\partial P}{\partial \hat{z}} \neq -\frac{\partial F_G}{\partial A_P}$, contrary to the Kelvin-Planck statement of the second law of thermodynamics (Equation 8).

In accordance with the discussion in Section 5.3, under the influence of the SND, $\mathcal{A}$ is $U - TS_{ne}$, not $U - TS_{eq}$. In general, $\frac{\partial S_{ne}}{\partial \hat{z}} \neq \frac{\partial S_{eq}}{\partial \hat{z}}$ and $\frac{\partial S_{ne}}{\partial A_P} \neq \frac{\partial S_{eq}}{\partial A_P}$. Therefore, as the $S_{eq}$-based Helmholtz free energy describes the equilibrium performance, the same formula of $\mathcal{A} = U - TS$ is incompatible with the nonequilibrium system. As long as $F'_{ij} \neq F'_{ji}$ (Inequality 11), no Massieu potential may be used as the counterpart of $\mathcal{A}$ for the entire system.

## 5.5 Sectional equilibrium and local Helmholtz free energy

As shown in Section 5.4, the conventional methods of equilibrium thermodynamics are no longer applicable to the nonequilibrium model system. Notice that the basic kinetic and dynamic theories can always be used (e.g., Newton's laws), which is how the governing equations of $P$ and $F_G$ (Equations 1 and 2) are obtained in Section 2.2.

Another approach to derive Equations (1) and (2) is to separately analyze the sectionally equilibrium domains for different thermodynamic forces, by using the local Helmholtz free energy. Each domain should be virtually isolated from the rest of the system, and contain only the chaotic and ergodic area, excluding the SND. The domain of $P$ is the plain, for which the local Helmholtz free energy is $\mathcal{A}_1 = N_P k_B T - TS_1$; the domain of $F_G$ is the plateau, for which the local Helmholtz free energy is $\mathcal{A}_2 = N_G k_B T + N_G mg\hat{z} - TS_2$. It can be seen that $-\frac{\partial \mathcal{A}_1}{\partial A_P} = \frac{N_P k_B T}{A_P} = P$, and $\frac{\partial \mathcal{A}_2}{\partial \hat{z}} = N_G mg = F_G$.

## 5.6 Topics of future study

In order to have a nontrivial nonequilibrium effect in Figure 1(B), the gravitational field must be ultra-strong, at the level of neutron stars or small black holes. In addition to gravity and pressure, other thermodynamic forces relevant to the SND include inertial force, Coulomb force, magnetic moment, angular momentum, entropy-related thermodynamic forces (e.g., osmotic pressure and degeneracy pressure), surface and interface tension, among others. Associated with the particle movement across a SND, there may be unconventional phenomena of mass and heat transfer. The working medium can be a gas, a Fermi gas, a plasma, a condensed matter, etc. It would be interesting to explore whether SND-type mechanisms may occur on the small time/length scales (e.g., superstring theory and quantum mechanics), in the middle ranges (e.g., molecular engineering, nanomaterials, nanofluidics, life science, etc.), on the large time/length scales (e.g., large-$g$ environments, the weak photon scattering in space, etc.), and in information theory. Moreover, the local nonchaoticity of the SND influences the basic assumptions of the Boltzmann equation and the H-theorem. These will be important topics in the future study.

## 6. Concluding Remarks

In the current research, we investigate the concept of spontaneously nonequilibrium dimension (SND). The model system contains a large number of billiard-like particles, randomly moving across a locally nonchaotic energy barrier (Figure 1B). The width of the barrier is much smaller than the mean free path of the particles.

The SND-based system does not obey the second law of thermodynamics. It can produce useful work in a cycle by absorbing heat from a single thermal reservoir. The second law of thermodynamics requires that the cross-influence of thermally correlated thermodynamic forces is symmetric (Equation 7). For the system under investigation, it dictates that the particles must follow the Maxwell-Boltzmann distribution (Equation 10). However, without extensive particle collision in the narrow barrier, there is no mechanism for the system to reach thermodynamic equilibrium. Our Monte Carlo simulation confirms that the particle distribution is indeed intrinsically nonequilibrium (Figure 5). As a result, Equation 7 is violated (Inequality 11). Remarkably, in an isothermal cycle, the produced work is significantly greater than the consumed work (Figure 6). Such a phenomenon cannot be explained in the conventional framework of statistical mechanics (e.g., Sections 5.4 and 5.5). The process does not involve Maxwell's demon. SND may have various configurations, e.g., with a switchable border or distributed nonchaotic traps (Figure 9).

**Appendix: Nomenclature**

$\mathcal{A}$: Helmholtz free energy

$\mathcal{A}_1$: Local Helmholtz free energy for the plain

$\mathcal{A}_2$: Local Helmholtz free energy for the plateau

$A_G$: Area of the upper plateau

$\tilde{A}_G = A_G \cdot \exp(-\hat{z}/z_0)$: Adjusted plateau area

$A_P$: Area of the lower plain

$A_{Pu}$ and $A_{Pl}$: Expanded and compressed plain areas, respectively

$d$: Particle diameter

$d_0$: Normalization factor of $d$

$D_G$: Diameter of the plateau

$\mathcal{E}_b$: Bias voltage

$F_G$: Support force of the plateau

$F_{G0} = mgN$: Normalization factor of $F_G$

$F_{Gc}$: Component of $F_G$ caused by the centrifugal force

$F_{Gg}$: Component of $F_G$ caused by the particle weight

$F_i$ and $F_j$: Thermodynamic forces

$F'_{ij} = \frac{\partial F_i}{\partial \tilde{x}_j}$: Cross-influence of $\tilde{x}_j$ on $F_i$

$F'_{ji} = \frac{\partial F_j}{\partial \tilde{x}_i}$: Cross-influence of $\tilde{x}_i$ on $F_j$

$g$: Gravitational acceleration

$g^*$: Normalization factor of $g$

$g_n$: Standard gravity

$i$ and $j$: Two large ergodic and chaotic areas

I, II, III, and IV: System states

$J_{ne}$: Flux of conduction electrons

$k_B$: Boltzmann constant

$K$: Particle kinetic energy

$\bar{K} = k_B T$: Average particle kinetic energy

$\bar{K}_0$: Characteristic $\bar{K}$ in the reference system

$K_z$: z-direction particle kinetic energy

$L_0$: System size

$L_G$: Circumference of the plateau

$\hat{L}$: Width of the chaotic ramp

$m$: Particle mass

$n_t$: Number of the data points

$N$: Total particle number

$N_G$: Number of the particles on the plateau

$\bar{N}_G = N_G/N$

$N_P$: Number of the particles on the plain

$N_{\mu\nu}$: Difference in $N_G$ between the $\mu$-th microstate and the $\nu$-th microstate

$p$: Probability density of the particle velocity

$p_z$: Probability density of $v_z$

$p_\mu$ and $p_\nu$: Probabilities of the $\mu$-th microstate and the $\nu$-th microstate, respectively

$P$: In-plane pressure of the plain

$P_0 = N\bar{K}/A_G$: Normalization factor of $P$

$P_G$: In-plane pressure of the plateau

$s_t$: Standard deviation

$S$: Entropy

$S_1$: Entropy of the plain

$S_2$: Entropy of the plateau

$S_{eq}$: Maximum entropy at thermodynamic equilibrium

$S_{ne}$: Nonequilibrium maximum entropy

$S_Q$: Maximum possible entropy of steady state

$\tilde{t}_p$: Characteristic time for a particle to change direction in the transition step

$\bar{t}_C$: Characteristic interval between particle collisions

$\bar{t}_T$: Characteristic travel time

$T$: Temperature

$U$: Internal energy

$v$: Particle velocity

$\bar{v}$: Average particle velocity

$\bar{v}_{\text{rms}}$: Root mean square particle velocity

$\bar{v}_x$: Average particle velocity in one direction

$v_z$: z-component of particle velocity

$\bar{v}_z$: Average $v_z$

$W_+$ and $W_-$: Works done and received by the plain, respectively

$W_G$: Work consumed by $F_G$

$W_i$ and $W_j$: Works produced/consumed by $F_i$ and $F_j$, respectively

$W_P$: Work produced by $P$

$W_{\text{tot}}$: Overall produced work

$x$: Horizontal dimension

$\tilde{x}_i$ and $\tilde{x}_j$: Conjugate variables of $F_i$ and $F_j$, respectively

$y$: Horizontal dimension

$z$: Vertical dimension

$\hat{z}$: Height of the transition step or vertical plane

$z_0 = \bar{K}/(mg)$: Characteristic step height

$\hat{z}_0 = 20$: Normalization factor of $\hat{z}$

$Z$: Partition function

$\hat{\alpha}$: Parameter of the non-Boltzmann distribution

$\beta = \frac{1}{k_B T}$: Thermodynamic beta

$\delta = \delta_{ji}/\delta_{ij}$: Crossing ratio

$\delta_0 = e^{-\beta mg\hat{}}$: Boltzmann factor

$\delta_{ij}$: Probability for the particles to move from area $i$ to $j$

$\delta_{ji}$: Probability for the particles to move from area $j$ to $i$

$\tilde{\Delta}(\bullet)$: Difference between the equilibrium state and the nonequilibrium steady state

$\Delta E_p$: Energy barrier of the pawl

$\Delta \bar{p}_x$: Change in momentum along the normal direction

$\Delta q_i$: Heat loss caused by $F_i$

$\Delta q_j$: Heat absorption caused by $F_j$

$\Delta t_0$: Timestep in the computer simulation

$\Delta \bar{t}$: Number of timesteps for the calculation of $P$

$\kappa_{\mu\nu}$: Probability ratio between the $\mu$-th microstate and the $\nu$-th microstate

$\lambda_F$: Nominal mean free path of the particles

$\hat{\rho}$: Particle density ratio

$\bar{\hat{\rho}}$: Average value of $\hat{\rho}$

$\rho_0$: Particle density at the bottom of the plane

$\rho_G = N_G/A_G$: Particle density on the plateau

$\rho_i$ and $\rho_j$: Steady-state particle densities in area $i$ and area $j$, respectively

$\hat{\rho}_I$: $\hat{\rho}$ associated with the lowered plateau

$\hat{\rho}_{II}$: $\hat{\rho}$ associated with the raised plateau

$\rho_P = N_P/A_P$: Particle density in the plain

$\sigma$ and $\tilde{\sigma}$: Constant terms for the entropy analysis